\documentclass[12pt]{article}

\usepackage{cite}
\usepackage{epsfig}
\usepackage{amsmath}
\usepackage{array}
\usepackage{bbold}

\def\mathswitch#1{\relax\ifmmode#1\else$#1$\fi}
\def\mathswitchr#1{\relax\ifmmode{\mathrm{#1}}\else$\mathrm{#1}$\fi}

\newcommand{\tev}{\,\, \mathrm{TeV}}
\newcommand{\gev}{\,\, \mathrm{GeV}}
\newcommand{\mev}{\,\, \mathrm{MeV}}

\newcommand{\eps}{\epsilon}
\newcommand{\tr}{{\rm tr}}

\newcommand{\mycaption}[1]{\caption{\sl #1}}

\begin{document}
\thispagestyle{empty}

\def\thefootnote{\fnsymbol{footnote}}

\begin{flushright}
ZU-TH 14/10\\
ANL-HEP-PR-10-54
\end{flushright}

\vspace{1cm}

\begin{center}

{\Large\sc {\bf Multi-Photon Signals from Composite Models at LHC}}
\\[3.5em]
{\large\sc
A.~Freitas$^1$, P.~Schwaller$^{2,3,4}$
}

\vspace*{1cm}

{\sl $^1$
Department of Physics \& Astronomy, University of Pittsburgh,\\
3941 O'Hara St, Pittsburgh, PA 15260, USA
}
\\[1em]
{\sl $^2$
Institut f\"ur Theoretische Physik,
        Universit\"at Z\"urich, \\ Winterthurerstrasse 190, CH-8057
        Z\"urich, Switzerland
}\\[1em]
{\sl $^3$
HEP Division, Argonne National Laboratory,\\
 9700 Cass Ave, Argonne, IL 60439, USA}
\\[1em]
{\sl $^4$
Department of Physics, University of Illinois, \\845 W Taylor St, Chicago, IL 60607, USA}

\end{center}

\vspace*{2.5cm}

\begin{abstract}
We analyze the collider signals of composite scalars that emerge in 
certain little Higgs models and models of vectorlike confinement. Similar 
to the decay of the pion into photon pairs, these scalars mainly decay 
through anomaly-induced interactions into 
electroweak gauge bosons, leading to a distinct signal with three or more 
photons in the final state. We study the standard model backgrounds for these signals,
and find that the LHC can discover these models over a large range of 
parameter space with 30~fb$^{-1}$ at 14 TeV. An early discovery at the current 7 TeV run is possible in some regions of parameter space. We also discuss possibilities to measure the spin of the particles in the $\gamma \gamma$ and $Z \gamma $ decay channels. 
\end{abstract}

\def\thefootnote{\arabic{footnote}}
\setcounter{page}{0}
\setcounter{footnote}{0}

\newpage

\section{Introduction}

New physics models that involve new strong dynamics often manifest themselves at low energies through new scalar or pseudoscalar fields. Examples of such models are Technicolor theories \cite{tcref} as well as composite Higgs and little Higgs models \cite{schmaltz}. Recently another class of models that are not directly involved in electroweak symmetry breaking has emerged, going by the name of vector-like confinement \cite{vec1}. 

The lowest order effective Lagrangians that describe these models often have an additional symmetry that forbids the decay of the lightest pseudoscalar at the tree level. The most prominent example for such a behavior is the low energy effective theory of QCD, where the decay of the neutral pion is only understood after including a higher order term, the Wess-Zumino-Witten (WZW) term \cite{WZW}, into the effective action. The same happens in little Higgs models, where T-parity \cite{LHT} is only broken after the WZW term is added \cite{hillsq,lhtwzw}. 
More generally, when the fermions that condense to form the pseudoscalars come in vector-like representations of all gauge interactions, the most important decay channel for the lightest scalar arises from the WZW term. 

The nonvanishing three-point interactions in the WZW term contain at least two gauge bosons, therefore the pseudoscalar will decay dominantly into the lightest available gauge bosons which are usually those of the standard model. Depending on the quantum numbers of the scalars, the decays can be into photons pairs or into $V \gamma$ or $VV'$, where $V$ and $V'$ can be any other standard model gauge bosons, including gluons \cite{Bai1}. 

In this work, we focus on pseudoscalar electroweak triplets that are pair produced at hadron colliders. These scalars have significant branching fractions into photon pairs, giving rise to signals with three or four photons in the final state. 

Multi-photon final states of this kind provide a promising signature at hadron colliders. One reason is that the standard model backgrounds for these processes are relatively low, so a signal can be found early even if the production cross section is at the femtobarn level, which is quite common for uncolored new states in BSM scenarios. Furthermore, owing to the energy resolution of the electromagnetic calorimeters, the mass of particles decaying to photon pairs can be determined with high accuracy. Finally, information about the spin of the particle can be obtained from angular distributions.

This paper is organized as follows: In the following section a brief overview is given about models that predict triplet scalars and their width and branching fractions are discussed. In section \ref{pheno:sigbg} signals and backgrounds are calculated for the 7 TeV and the 14 TeV LHC. In section \ref{pheno:mass} we determine how precise the mass and spin of the particle can be measured, before we summarize in section \ref{pheno:end}. 
\section{Pseudoscalar Triplets in Standard Model Extensions}
Our main objects of interest are pseudoscalar electroweak triplets $\phi_a = (\phi_a^+,\phi_a^0,\phi_a^-)$ that transform as $(3,0)$ under the electroweak $\rm SU(2) \times U(1)$ gauge symmetry. The leading interactions with the standard model are contained in the kinetic term
\begin{align}\label{eq:ekin:phia}
    {\cal L}_{\rm kin} = \tr \left( D^\mu \phi_a D_\mu \phi_a^\dagger \right) + {\cal O}(\phi_a^4)\,.
\end{align}
A light pseudoscalar $\phi_a$ can emerge as a pseudo-Goldstone boson when it is part of a larger multiplet of some global symmetry that is involved with electroweak symmetry breaking. In this case the above interactions will receive corrections suppressed by powers of $(v/f)^2$, where $v$ is the Higgs vacuum expectation value and $f$ is the breaking scale of the global symmetry. The scale $f$ typically lies around the TeV scale so we expect these corrections to be at most 10\%. Also the higher point interactions contained in \eqref{eq:ekin:phia} are not of interest here. 

We are in particular interested in cases where the $\phi_a^{\pm,0}$ are the lightest particles that are odd under an approximate parity symmetry. In moose models, such as the little Higgs model with X-parity \cite{Krohn:2008ye,mmx}, this happens due to the symmetry structure of the coset spaces, and is a remnant of the original T-parity in these models \cite{LHT}. 

While it is often assumed that little Higgs models originate from a theory that becomes strongly coupled around the 10 TeV scale, this UV completion is usually not specified. A concrete example of high-scale strongly interacting dynamics with a remnant parity is vector-like confinement \cite{vec1,vec2}. The benchmark model discussed in \cite{vec2} contains a pseudoscalar triplet with the required quantum numbers. In that case the accidental parity symmetry appears because the fermions that condense to yield the pseudo-Goldstone bosons transform in vector-like representations of the standard model gauge group. Similar models that also contain a pseudoscalar triplet were presented in \cite{Bai1,Bai2}. 

\begin{figure}
\begin{center}
\psfig{figure=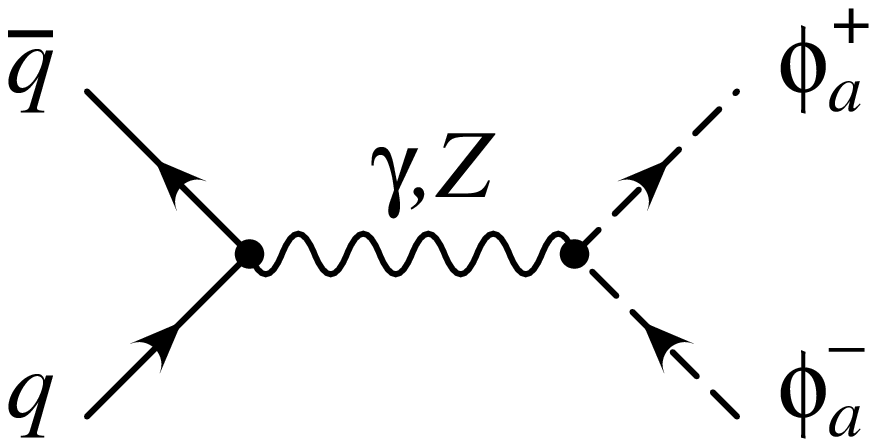, width=5cm}\hspace*{1cm}
\psfig{figure=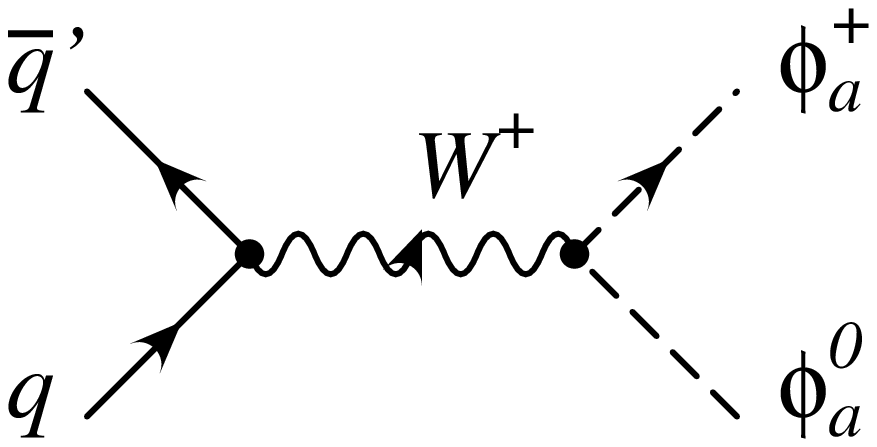, width=5cm}
\end{center}
\vspace{-1em}
\mycaption{Pair production diagrams for the scalar triplets at hadron colliders. The charge conjugate of the second diagram is not shown.}
\label{fig:phiprod}
\end{figure}
To a large extent the properties of the triplets are fixed by gauge invariance and by the parity symmetry, so they can be studied independently of the model they belong to. Masses for the triplet are generated through radiative corrections. The neutral and charged components of the triplet are expected to have roughly the same mass $m_a$, since a large mass splitting is constrained by the electroweak $T$ parameter.
The little Higgs model with X-parity \cite{mmx} predicts that mass to be of order of the electroweak scale. The same mass range is assumed in \cite{vec2}. For the present analysis, we will consider
\begin{align}
    100 \gev < m_a < 600 \gev\,.
\end{align}
Due to the approximate parity symmetry, the triplets are mostly produced in pairs via an intermediate $W^\pm$ or $Z$ boson, through interactions contained in the kinetic term (\ref{eq:ekin:phia}). The relevant Feynman diagrams for production at hadron colliders are shown in figure \ref{fig:phiprod}. Note that there is no direct  $\phi_a^0\phi_a^0$ production
due to Bose symmetry. 

Additional contributions to $\phi_a$ pair production from BSM particles are possible. For example, the model in \cite{vec2} contains an additional vector boson that is produced in the $s$-channel and decays into $\phi_a$ pairs. The cross sections obtained from the diagrams in figure \ref{fig:phiprod} can nevertheless be used as minimal expectations for the pair production rates. 

Since the lowest order Lagrangian is symmetric under pseudoscalar parity the $\phi_a$ can only decay through the fourth order WZW term:
\begin{align}\label{eq:wzw}
    \Gamma_{\rm WZW} & = \frac{N}{16 \pi^2 f} \int d^4x \, \eps_{\mu\nu\rho\sigma} 
    \,\tr\left(\phi_a F_{\mu\nu} F_{\rho\sigma} \right) + \dots \,, 
\end{align}
where the dots denote terms that do not contribute to $\phi_a$ decays at leading order, and the gauge couplings have been absorbed into the definition of the field strength tensor. Note that unlike the case of vector boson decays here the neutral $\phi^0_a$ can decay into two photons. On the other hand the decay $\phi_a^0 \rightarrow W^+ W^-$ is not possible since the corresponding anomaly coefficient vanishes. 

\begin{figure}
\begin{center}
\epsfig{figure=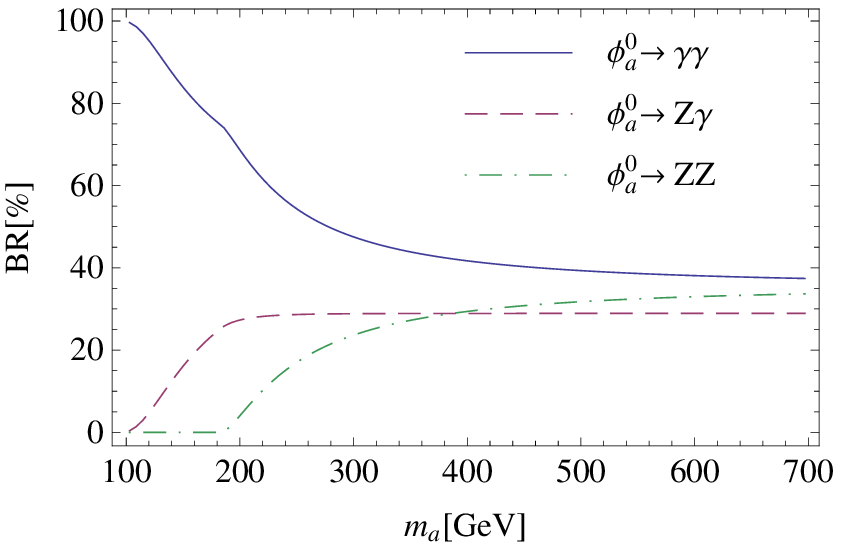, width=7.5cm}%
\end{center}
\vspace{-1.5em}
\mycaption{Branching fractions of $\phi_a^0$ depending on the mass $m_a$. 
}
\label{fig:phi0br}
\end{figure}

The relative branching fractions between $\phi^0_a$ decay modes are completely fixed by this expression\footnote{The branching fractions for WZW decays of $\phi_a$ were already presented in the arXiv version of \cite{proc09}.}---the nontrivial dependence on $m_a$ that is displayed in figure \ref{fig:phi0br} is purely from kinematic suppression. The total widths in addition depend on the integer $N$ and on the scale $f$, and typically lie in the $\rm keV$ range. 

\begin{figure}
\begin{center}
\epsfig{figure=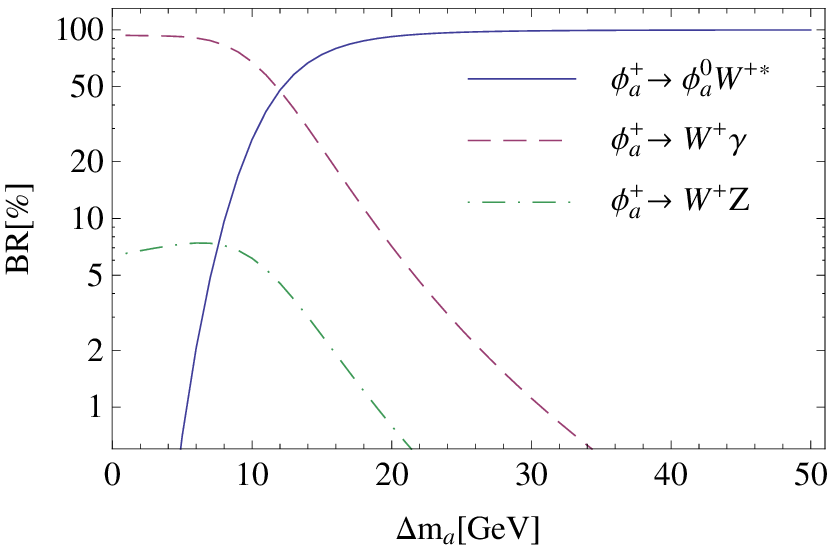, width=7.5cm}%
\end{center}
\vspace{-1.5em}
\mycaption{Branching fractions of $\phi_a^+$ as a function of the mass splitting $\Delta m_a$, for $m_a=300 \gev$.}
\label{fig:phipbr}
\end{figure}

After electroweak symmetry breaking, radiative corrections introduce an additional splitting between the neutral and the charged components of the multiplet. For purely electromagnetic interactions, and in the limit $m_a \gg m_W$, it was found in \cite{Cirelli} that 
\begin{align}
    \Delta m_a = m_{\phi_a^+} - m_{\phi_a^0} \approx 170 \mev\,.
    \label{eq:masssplit}
\end{align}
Since there are additional contributions to this mass splitting in models with extended gauge and scalar sectors, we treat $\Delta m_a$ as a free parameter. The decay $\phi_a^\pm \rightarrow \phi_a^0 W^{\pm,*}$ becomes relevant around $\Delta m_a = 5 \gev$ and dominant for larger splittings, as shown in figure \ref{fig:phipbr}. This behavior is mostly independent of the absolute mass scale $m_a$.  
For small mass splitting the dominant decay mode for the charged triplet is $\phi_a^\pm \rightarrow \gamma W^\pm$ with a branching fraction decreasing from almost 100\% at $m_a=100\gev$ to around 80\% at $m_a = 700 \gev$. 
\vfill
\section{Signals and Backgrounds}
\label{pheno:sigbg}
\subsection{The signal}
\begin{figure}
\begin{center}
\epsfig{figure=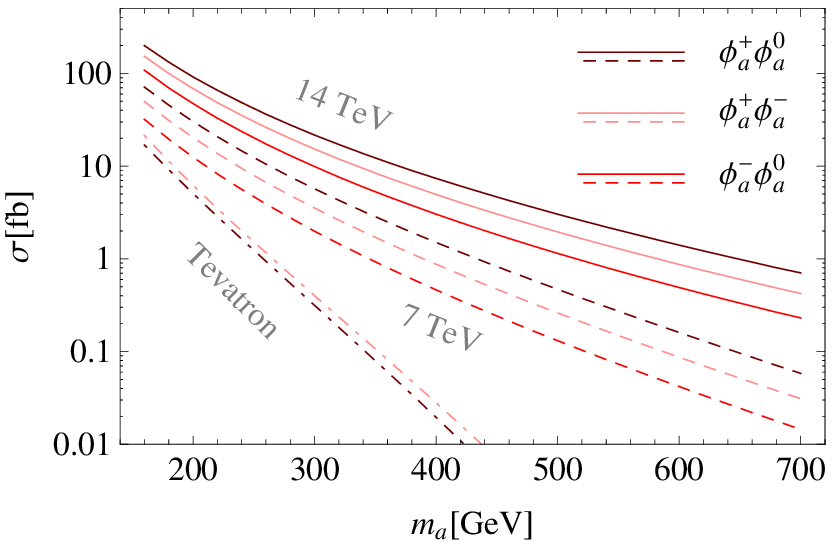, width=8cm}%
\end{center}
\vspace{-1.5em}
\mycaption{Production of $\phi_a\phi_a$ pairs at the 7 TeV LHC (dashed lines) and the 14 TeV LHC (solid lines). For comparison, the Tevatron production cross sections for $\phi_a^\pm \phi_a^0$ pairs (dark red dash-dotted) and $\phi_a^+ \phi_a^-$ pairs (light red dash-dotted) are also shown.}
\label{fig:lhcprod}
\end{figure}

The cross sections for $\phi_a^\pm \phi_a^0$ and $\phi_a^+\phi_a^-$ production are shown in figure \ref{fig:lhcprod} for the LHC running at $7\tev$ and $14\tev$ center of mass energy. With $14 \tev$ the production cross sections are sizable up to $m_a = 700 \gev$, while the reach of the current LHC run with $\sqrt{s} = 7\tev$ and 1 fb$^{-1}$ of target luminosity is clearly limited. 

In the case of a small mass splitting, $\Delta m_a \ll 5\gev$, the most interesting signal arises from $\phi_a^\pm \phi_a^0$ decaying into $3\gamma+W^\pm$. On the other hand, if the splitting is larger than about 5 GeV all production channels contribute to a $4\gamma + X$ signal \cite{proc09}. 

Our analysis will focus on the inclusive $3\gamma$ signal, $i.\,e.$ on the case of a small mass splitting. To be definite we will consider three scenarios with $m_a = 200$ GeV, 400 GeV and 600 GeV. We use CompHEP 4.5.1~\cite{comphep} with CTEQ6L1 parton distributions to simulate the processes
\begin{align}
    pp \longrightarrow \phi_a^\pm \phi_a^0 \longrightarrow \gamma\gamma\gamma W^\pm
\end{align}
at $\sqrt{s} = 7\tev$ and $14 \tev$. The resulting events are passed to PYTHIA 6.4 \cite{pythia} for hadronization and $W$ boson decay and through PGS4 \cite{pgs} with the CMS parameter set to account for detector efficiencies and energy smearing. The cuts used for event generation are summarized in table \ref{tab:cuts0} and the resulting cross sections for the 7 TeV and the 14 TeV LHC are given in table \ref{tab:sig0}.

\begin{table}
\begin{center}
\begin{tabular}{|c|c|}\hline
$p_T$ all photons & $>40$GeV \\ 
$|\eta|$ all photons & $<2.5$ \\ 
$\Delta R$ all photons & $>0.3$ \\ \hline  
\end{tabular}
\end{center}
\vspace*{-1em}
\mycaption{Cuts used to generate $pp \longrightarrow \phi_a^\pm \phi_a^0 \longrightarrow \gamma\gamma\gamma W^\pm$ events. We also require all photons to be separated from the $W^\pm$ by more than $0.3 \text{ rad}$. }
\label{tab:cuts0}
\end{table}
\begin{table}
\begin{center}
\begin{tabular}{|c|c|c|c|}\hline
    \textbf{LHC Energy $\backslash$ Mass} & \textbf{200 GeV} & \textbf{400 GeV} & \textbf{600 GeV} \\ \hline
    $7 \tev $& 17.3  & --- & --- \\\hline
    $ 14 \tev$ & 43.5  & 2.69 & 0.13 \\ \hline
\end{tabular}
\end{center}
\vspace*{-1em}
\mycaption{Detector level cross sections in fb for the $3\gamma+X$ signal for the benchmark masses. The cuts imposed on the photons are given in table \ref{tab:cuts0}. Detector efficiencies are estimated using PGS4. The 7~TeV run does not give a detectable signal for the heavier scenarios.}
\label{tab:sig0}
\end{table}

The standard model background to three photon production is of the same order as the signal, and will be discussed in the next sections. The backgrounds for $4\gamma$ signals are suppressed by an additional power of $\alpha$ and thus negligible. We will comment on the $4\gamma$ signal at the end of section \ref{lh3:lhcsen}.

Before turning to the backgrounds, let us briefly discuss the experimental bounds on the $3\gamma$ signal and on the mass $m_a$. 
The D\O\ experiment analyzed $3\gamma+X$ events in the search for fermiophobic Higgs bosons \cite{dzero0} in a sample corresponding to $0.83 \;{\rm fb}^{-1}$ of collected data. The absence of an excess of events in the sample translates into an upper bound on the production cross section of fermiophobic Higgs bosons $\sigma \leq 25.3 \;{\rm fb}$. For our model this is satisfied provided that $m_a \geq 150 \gev$. 

Another possibility for the model to show up at the Tevatron experiments is through a bump in the di-photon invariant mass spectrum. The most recent searches \cite{Abazov:2010xh,cdfpublic} for graviton resonances in that channel with $5.4\; {\rm fb}^{-1}$ of integrated luminosity however do not impose further constraints on $m_a$.

\subsection{Real Backgrounds}

There are two types of standard model backgrounds for the $3\gamma+X$ signal: real backgrounds with three or more photons in the final state, and fake backgrounds where one or more photons are actually jets that were misidentified in the detector. At the tree level pure photon final states are only produced from quark anti-quark initial states. Only at the NLO level the gluon gluon channels become available through a quark loop. This leads to large k-factors, in particular at the LHC where gluons are abundant. 

The main source of real backgrounds is direct three-photon production accompanied with any number of jets. This background has been generated using MadGraph/MadEvent \cite{madgraph} with the same cuts as for the signal, and was then processed through PYTHIA and PGS4 for initial and final state radiation and for modeling of detector effects. The radiative corrections to three-photon production are yet unknown, but based on experience from two-photon production \cite{2phot} we expect a large k-factor and therefore conservatively multiply this background by a factor of two.

The production of three photons together with a $W$ boson has a tiny cross section in the standard model and can be neglected. The background rates for the 7 TeV and 14 TeV LHC are given in table \ref{tab:bg0}.

\begin{table}
\begin{center}
\begin{tabular}{|c|ccc|ccc|}\hline
    \textbf{process} & \textbf{7 TeV:}&\textbf{ MG/ME} & \textbf{PGS4} & \textbf{14 TeV:}& \textbf{MG/ME} & \textbf{PGS4}  \\ \hline
    $3 \gamma + n \text{ jets}$ && 2.51  & 2.01  && 5.44  & 4.54 \\\hline
    $3\gamma + W^\pm$ && 0.0051 & 0.0036 && $0.014$ & $0.009$ \\ \hline
    $2 \gamma + n \text{ jets}$ && 7190  & 5.9 && 13700 & 8.9 \\ \hline 
\end{tabular}
\end{center}
\vspace*{-1em}
\mycaption{Real and fake backgrounds in fb for inclusive $3\gamma$ searches with $p_{T,\gamma}>40 \gev$ at LHC with $E_{\rm cm} = 7 \tev$ and $14 \tev$. The first column gives the partonic cross sections obtained with MadGraph/MadEvent, while the second column gives the cross section for the fraction of events that are reconstructed as $3\gamma + X$ events in PGS4. For the fake backgrounds the first column shows the cross section for $2\gamma + n \text{ jet}$ processes with $p_{T,\gamma}>40\gev.$ Backgrounds are multiplied by  two to account for NLO k-factors and uncertainties from matching.}
\label{tab:bg0}
\end{table}

\subsection{Fake Backgrounds}
The most important source of fake backgrounds are the processes $pp \rightarrow \gamma\gamma + n \text{ jets}$ where one of the jets is misidentified as a photon.

We use MLM type matching with the $k_T$ jet algorithm to match events with $\gamma\gamma + 0,1,2 \text{ jets}$ between MadEvent and PYTHIA with a matching scale of $30\gev$ \cite{madmatch}. The matched sample is then run through PGS4 to model detector effects and in particular to get an estimate for the fake jet rate and for the magnitude of the fake backgrounds. Table~\ref{tab:bg0} shows the resulting background after application of the cuts in table \ref{tab:cuts0} and multiplication by a k-factor of 2 to account for the cross section uncertainty due to leading order matching and NLO corrections. 

A few comments are in order. The event sample can contain additional photons from initial and final state radiation. Imposing the cut of $p_T>40\gev$ on all photons in the sample effectively removes most of this contribution. 

Additional hard photons in the final output are therefore due to jets that were reconstructed as photons in PGS4. Tests with a pure two-jet sample with $p_T > 40\gev$ show that the fake rate in PGS4 lies at the per mille level. This should be compared with the fake rates obtained by CMS \cite{cms_photon} using a full detector simulation. A fake photon candidate in the electromagnetic calorimeter is produced by about 1 in 200 jets. Using additional isolation cuts CMS obtains an additional rejection factor of 100 while keeping a reasonable photon efficiency (80\%), $i.\,e.$ the total rejection factor is 20000 corresponding to a fake rate of 0.005\%.\footnote{ATLAS \cite{atl_photon} gives a rejection rate of 7000 for $p_T>40\gev$.} This is about one order of magnitude better than the PGS4 estimate, at the expense of a reduced photon efficiency compared to PGS4 ($>90\%$).  

Considering that the optimized isolation cuts of Ref.~\cite{cms_photon} are not implemented in PGS4 its performance is acceptable, and in particular it is sufficient for the present work. Improving this performance would still be desirable, in particular since the fake backgrounds are larger than the real backgrounds.

\subsection{LHC Sensitivity}\label{lh3:lhcsen}

For masses $m_a$ above $400 \gev$ the backgrounds become comparable to the signal. It is therefore necessary to further increase the signal to background ratio while maintaining a large signal efficiency.

\begin{figure}
\begin{center}
\epsfig{figure=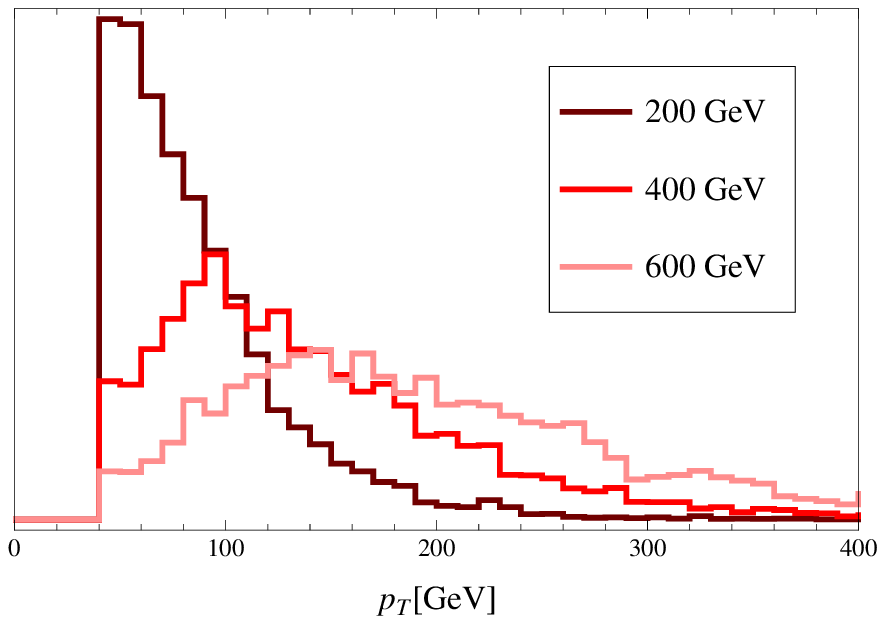, width=7.5cm}%
\epsfig{figure=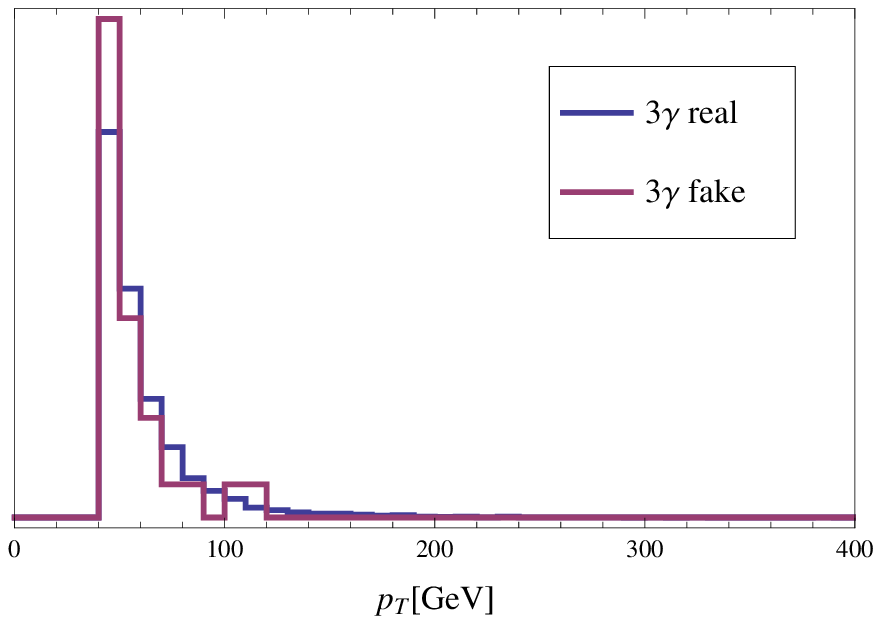, width=7.5cm}%
\end{center}
\vspace{-1.5em}
\mycaption{Transverse momentum distribution of the third hardest photon for signal (left) and background (right) events at the 14 TeV LHC. All distributions normalized to unity.}
\label{fig:ptdist}
\end{figure}

One possibility is to impose stronger cuts on the transverse momentum of the photons. Since the signal photons come from the decay of heavy particles their $p_T$ spectrum is relatively flat up to $p_T = m_a/2$ while the backgrounds fall off rather quickly with increasing $p_T$. The $p_T$ distribution of the third hardest photon for signal and background events is displayed in figure \ref{fig:ptdist}. 
The best results are obtained with a uniform $p_T$ cut on all photons. With a cut of $60 \gev$ ($80 \gev$) we obtain a background suppression of 66\% (86\%) while loosing at most 30\% (54\%) of the signal for $m_a=200\gev$, and much less for larger values of $m_a$. 

\begin{figure}
\begin{center}
\epsfig{figure=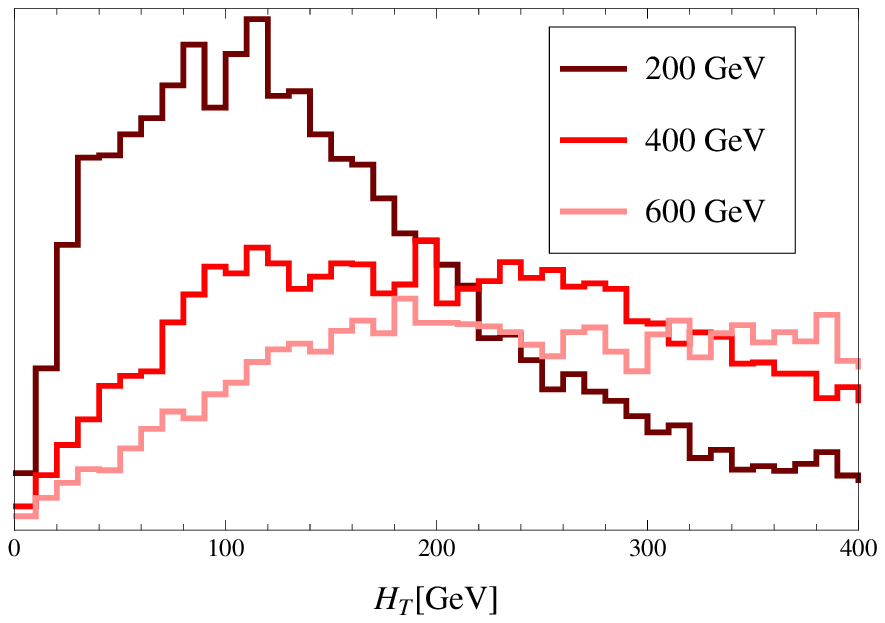, width=7.5cm}%
\epsfig{figure=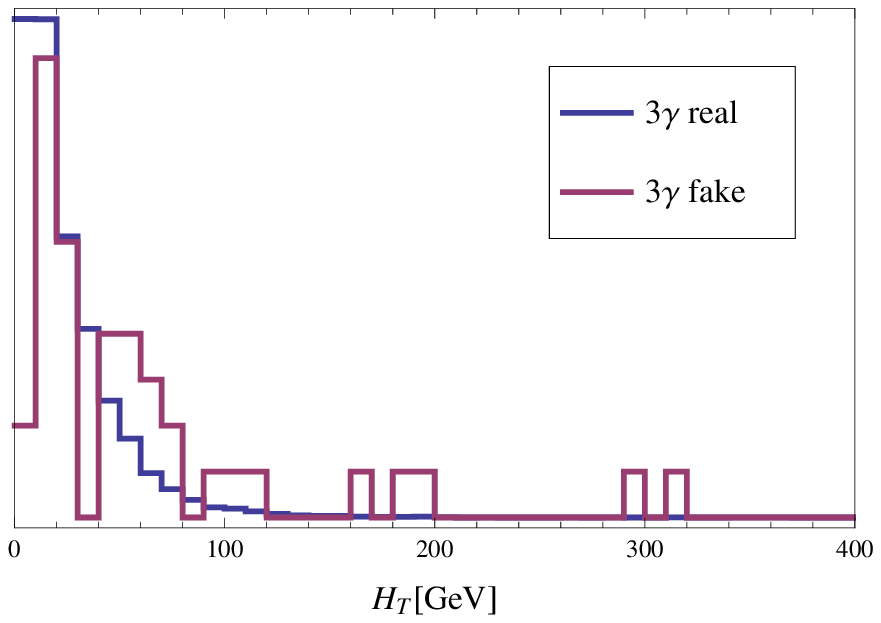, width=7.5cm}%
\end{center}
\vspace{-1.5em}
\mycaption{$H_T$ distributions for signal (left plot) and background (right plot) events at the LHC with $E_{\rm cm} = 14\tev$. The unsteady shape of the fake background distribution is due to the smallness of the fake sample. The results were cross checked against the $H_T$ distribution of two photons plus one jet and found to agree.}
\label{fig:htdist}
\end{figure}

In signal events the three photons effectively recoil against a W-boson which can carry away a significant amount of transverse momentum. On the other hand most background events will be balanced in transverse momentum except for those cases where an additional hard jet is present. 
We define 
\begin{align}
    H_T \equiv \sqrt{\left(\sum_{i=1}^3 p_x^i\right)^2 + \left(\sum_{i=1}^3 p_y^i\right)^2 }
\end{align}
as a measure of the $p_T$ imbalance of the three photon system. The $H_T$ distributions for signal and backgrounds are displayed in figure \ref{fig:htdist}, which shows that the backgrounds are a concentrated in the region $H_T<100\gev$. 

As final cuts for the analysis we choose $p_{T,\gamma}>80 \gev$ for the three hardest photons in the event and $H_{T}>80\gev$. The effects of these cuts on the signal and backgrounds are summarized in table \ref{tab:sbcuts}. They improve the signal to background ratio by more than an order of magnitude. 
\begin{table}
\begin{center}
\begin{tabular}{|l|ccc|c|}\hline
     & $p_T>80 \gev $&$ H_T>80 \gev $ & \textbf{combined} & \textbf{Events}  \\ \hline
    $200 \gev$ & 46.4\% & 75.3\% & 36.0\% & 470 \\
    $400 \gev$ & 80.8\% & 90.3\% &  72.7\% & 59 \\ 
    $600 \gev$ & 93.2\% & 94.1\% &  87.7\% & 3.4 \\ \hline 
    \text{real BG} & 13.6\% & 3.2\% & 1.3\% &  1.8 \\ 
    \text{fake BG} & 10.7\% & 20.5\% & 3.5\% & 9.3 \\ \hline 
\end{tabular}
\end{center}
\vspace{-1em}
\mycaption{Cut efficiencies on signal and background events. Shown are the fraction of events that pass the indicated cuts. The last column shows the expected number of Events at the LHC with $30 \;\rm{fb}^{-1}$ after the cuts have been applied.}
\label{tab:sbcuts}
\end{table}
Also shown in table \ref{tab:sbcuts} are the expected signal and background events at the LHC for a luminosity of $30 \; \rm{fb}^{-1}$. From the results one can conclude that a discovery with a statistical significance of 5 standard deviations is possible for values of $m_a$ up to about $500\gev$. 

For the heavy scenario with $m_a=600\gev$ a 5$\sigma$ discovery would only be possible with the full design luminosity of the LHC and further reduction of the backgrounds by imposing stronger cuts on the photon transverse momenta and $H_T$. Note that the signal selection could be improved further by trying to find a peak in the di-photon invariant mass spectrum above the smooth background. 

\vspace{\bigskipamount}

The LHC is currently running at a center of mass energy of 7 TeV and is supposed to collect at least $1 \;{\rm fb}^{-1}$ before the shutdown at the end of 2011. It is interesting to ask whether this can suffice to detect a signal at least in the light scenario ($m_a=200\gev$). From table \ref{tab:sig0} we expect 17 signal events. The backgrounds are smaller than for the 14 TeV case and have an even steeper $p_T$ spectrum. Imposing a $p_T$ cut of 60 GeV on all photons we expect 11.4 signal events and 2.0 background events with $1 \;{\rm fb}^{-1}$, thus a discovery at the $5\sigma$ level is possible. 

\vspace{\bigskipamount}

Finally let us comment on the case of a large mass splitting between $\phi^\pm_a$ and $\phi^0_a$, where the dominant signal is from the $4\gamma + X$ final state. In addition to the $\phi^\pm_a \phi_a^0$ channel now also the $\phi_a^+\phi_a^-$ channel contributes to the $4\gamma+X$ final states. 
Radiating an additional photon in any of the discussed backgrounds events costs an additional factor of $\alpha=1/128$, while requiring an additional fake photon reduces the cross section even further. One can therefore safely assume that the standard model backgrounds to the $4\gamma$ signal are negligible. 

\begin{figure}
\begin{center}
\epsfig{figure=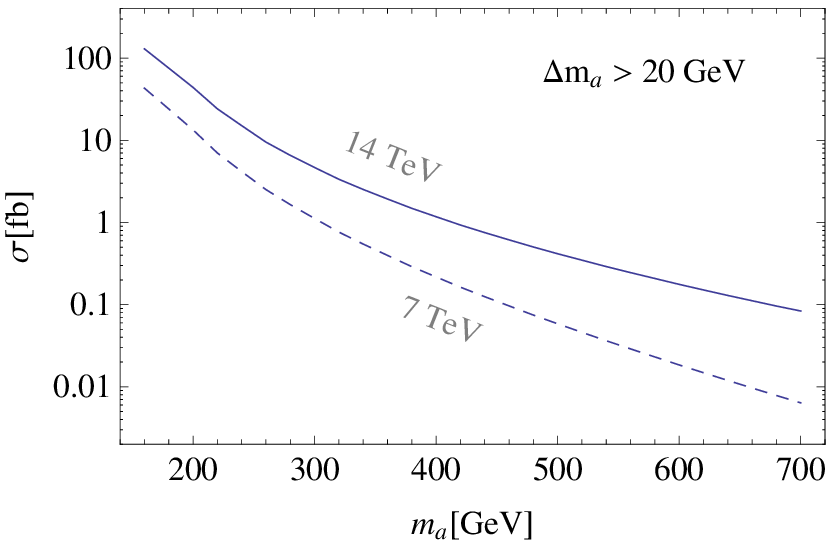, width=7.5cm}%
\end{center}
\vspace{-1.5em}
\mycaption{Cross sections for the $4\gamma+X$ signal for a large mass splitting, $\Delta m_a> 20\gev$.}
\label{fig:4A}
\end{figure}
The cross section for the $4\gamma+X$ signal in the large splitting case is determined by multiplying the $\phi_a$ pair production rates in figure \ref{fig:lhcprod} with the branching fractions for $\phi^\pm_a \rightarrow \phi^0_a W^{\pm*}$ and $\phi^0_a\rightarrow \gamma \gamma$. Assuming $\Delta m_a > 20 \gev$, almost all of the charged scalars decay into $\phi^0_a$ and an off-shell $W^\pm$, with BR$(\phi^\pm_a \rightarrow \phi^0_a W^{\pm*})>95\%$. From an explicit simulation, using as before CompHEP 4.5.1, PYTHIA 6.4, and PGS4 with the basic cuts in table~\ref{tab:bg0}, we obtain that 44\% of the signal events are reconstructed by the detector for $m_a=200\gev$. It is expected that the signal efficiency increases slightly for larger $m_a$, but for simplicity we will assume the same value for the entire mass range. The results are shown in figure \ref{fig:4A}. Since the backgrounds are negligible, we require a minimum of 5 signal events for a discovery. This gives a discovery range for the current 7~TeV LHC run up to $m_a= 250\gev$, while the LHC with $\sqrt{s}=14\tev$ will be able to probe and the full parameter range up to $m_a= 600\gev$ with about $30\; {\rm fb}^{-1}$. 
\section{Measuring Particle Properties}
\label{pheno:mass}
\subsection{Mass Measurement}
The mass of the neutral $\phi^0_a$ can be determined directly from the di-photon invariant mass distribution in the three photon samples. Since the width of $\phi^0_a$ is small the quality of the measurement is entirely determined by the energy resolution of the detector. 

\begin{figure}
\begin{center}
\epsfig{figure=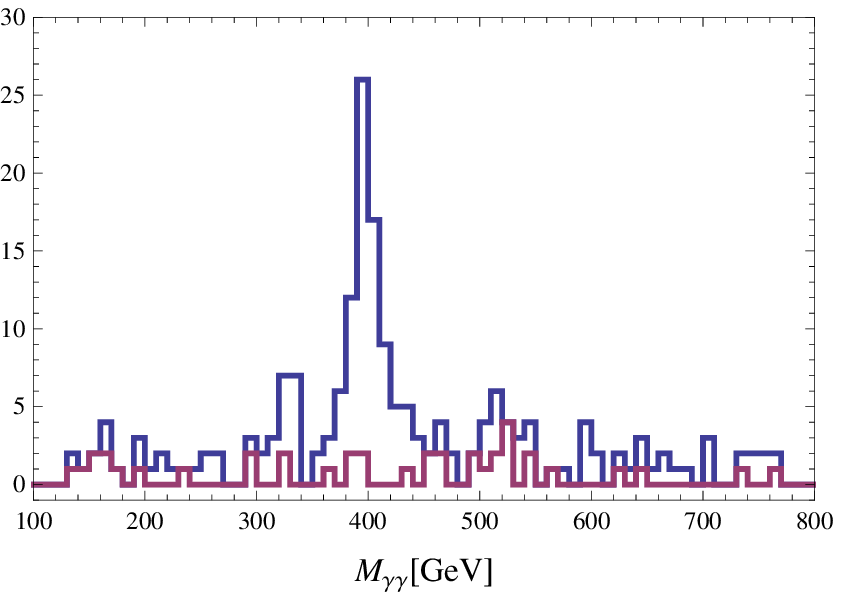, width=7cm}%
\hspace{1em}
\epsfig{figure=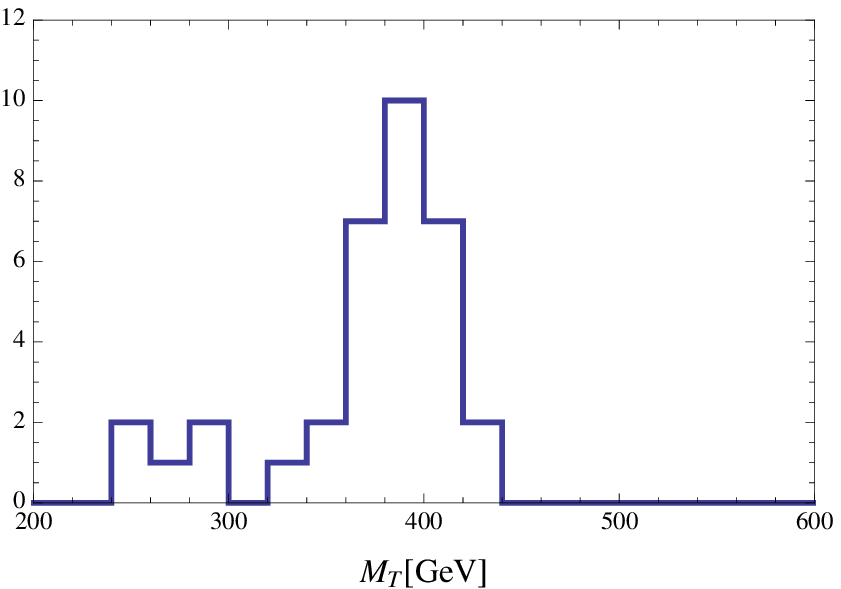, width=7cm}%
\end{center}
\vspace{-1.5em}
\mycaption{Left: Invariant mass distribution of photon pairs in signal plus background events (blue) and backgrounds only (red), for $30 \;{\rm fb}^{-1}$. Note that each event gives three entries, one for each possible pairing of photons. 
Right: Transverse mass distribution from  $\phi^\pm_a \rightarrow W^\pm \gamma \rightarrow \gamma \ell^\pm {E\!\!\!\slash}_T$ decays, with events corresponding to $100 \;{\rm fb}^{-1}$ of collected data at LHC. Both plots correspond to $m_a=400\gev$.
}
\label{fig:mass0}
\end{figure}

The left plot in figure \ref{fig:mass0} shows the invariant mass distribution of photon pairs from signal and background events for $m_a = 400 \gev$ and a luminosity of $30 \; \rm{fb}^{-1}$, corresponding to 59 signal and 11 background events. 
A sharp peak around $m_{\gamma \gamma} = 400 \gev$ is clearly visible on top of the standard model background and combinatoric background from the signal.

Measuring the mass of the charged $\phi^\pm_a$ is more difficult since it decays into $\gamma W^\pm$. Decays of the $W$ boson into electrons or muons, either directly or via taus that decay leptonically, occur in about $25\%$ of all cases and lead to a $3\gamma + \ell$ signal that is essentially free of backgrounds. We also include the taus from hadronic decays that are reconstructed by PGS. A fraction of the $W$ boson momentum is carried away by one or several neutrinos.
Since neutrinos are the only physical source of missing energy and they are essentially massless, the $\phi^\pm_a$ mass can be determined accurately from the transverse mass
\begin{align}
	M_T^2 = m_{\ell \gamma}^2 + 2 ( e_{\ell\gamma} e_{\rm miss} - \mathbf{p}_{T,\ell\gamma}\cdot \mathbf{p}_{T,\rm miss}),
\end{align}
where $m_{\ell \gamma}$ is the invariant mass of the lepton-photon system,
$e_{\ell \gamma} = \sqrt{m_{\ell\gamma}^2 + p_{T,\ell\gamma}^2}\;$, $e_{\rm miss} = |\mathbf{p}_{T,\rm miss}|$  and $\mathbf{p}_{T}$ denotes the transverse momentum of the indicated particle set. 
To obtain the transverse mass distribution we identify the photon pair from the $\phi^0_a$ decay by the requirement that $m_{\gamma\gamma}$ agrees with $m_{\phi^0_a}$ within 5\%, and then combine the remaining photon with the lepton, which we require to have $p_{T,\ell} > 30 \gev$. We further require that $|\mathbf{p}_{T,\rm miss}|>30 \gev$ to reduce the background from misidentified leptons. Events where no photon pair can be identified uniquely by the $m_{\gamma\gamma}$ window condition are discarded. 

The resulting transverse mass distribution is shown in figure \ref{fig:mass0}. For this analysis we use a sample with 200 signal events corresponding to $100 \;{\rm fb}^{-1}$, whereof 43 events satisfy our cuts. In spite of the small number of events the endpoint of the $M_T$ distribution near $m_{\phi^+_a} = 400 \gev$ is clearly visible. Theoretically one would expect a sharp cutoff at the endpoint, which however is washed out due to detector smearing, in particular the mis-measurement of the missing transverse momentum. 
\subsection{Spin and CP Properties}
Some information about the spin of the resonance can already be extracted from the nature of the final state. A two-photon resonance can originate only from a spin-zero of spin-two particle [we do not consider spins larger than two].

To distinguish further between the spin-zero and spin-two cases, one has to resort to angular distributions involving the final-state photons.
An observable that is sensitive to the spin of $\phi_a^0$ is the distribution with respect to the angle $\Phi$ between the production plane and decay plane of $\phi_a^0$. Once the mass of the resonance has been measured, this angle can be constructed from the momenta of the two photons that reconstruct the resonance together with the beam axis. 
Assuming an unpolarized initial state a nontrivial distribution can only arise from interference between different helicity states and would therefore rule out a spin-zero interpretation of the signal \cite{Murayama:2009jz}. 

\begin{figure}
\begin{center}
\epsfig{figure=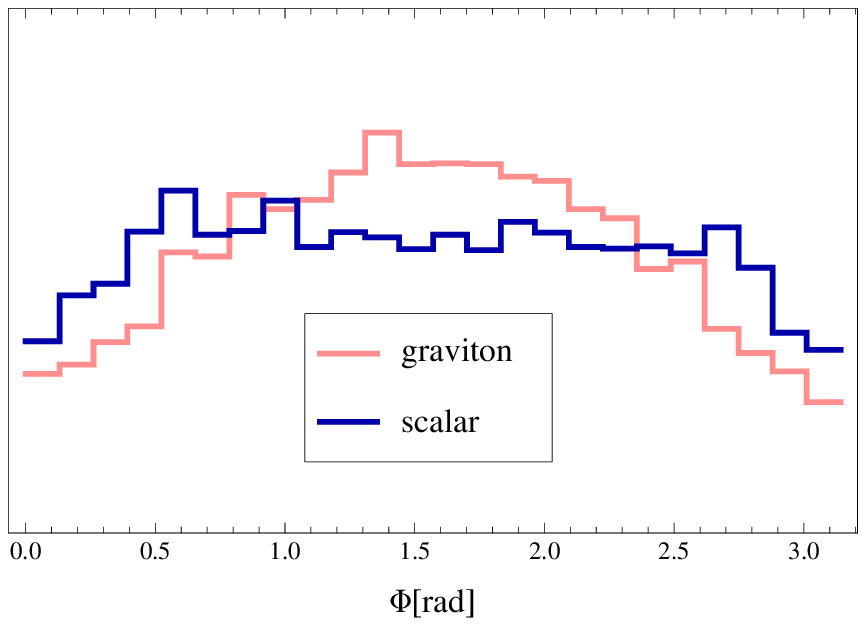, width=8.5cm}%
\end{center}
\vspace{-1.5em}
\mycaption{Normalized distribution of the angle $\Phi$ between the production and the decay plane of $\phi_a^0$ (dark blue).  For comparison, the same distribution is shown for a graviton decaying into two photons in the process $pp \rightarrow G \gamma \rightarrow 3 \gamma$ (light red). Both distributions use the cuts of table \ref{tab:cuts0}, and $m_{\phi_a^0} = m_G = 400 \gev$.}
\label{fig:phidist}
\end{figure}

The distribution of $\Phi$ in a sample of 10,000 $3\gamma$ events with $m_{\phi_a^0} = 400 \gev$ is shown in figure \ref{fig:phidist}. While one would expect this distribution to be completely flat, some angular dependence is introduced by the selection cuts, which explains the lower counting rates around $\Phi=0$ and $\Phi=\pi$. 
For comparison, we show the same distribution for a KK-graviton $G$ decaying into two photons in the process $pp \rightarrow G\gamma \rightarrow 3 \gamma$, with $m_G = 400 \gev$. This distribution was produced using an implementation of massive spin-two particles into MadGraph/MadEvent \cite{Hagiwara:2008jb}.

To estimate how well the two cases can be distinguished with a realistic number of 600 events (corresponding to 300~fb$^{-1}$ integrated luminosity), we calculate the ratio of the number of events in the interval $\pi/4 < \Phi < 3 \pi/4$ divided by the number of events lying outside of this interval. For the case of the pseudoscalar resonance $\phi_a^0$ we obtain $1.14 \pm 0.09$ whereas we get $1.74\pm 0.15$ for the graviton resonance. The central value and errors have been obtained by randomly picking 600 events from the full samples multiple times and extracting the mean and standard deviation. It is therefore possible to distinguish the two cases considered here at the three sigma level.

One drawback of this method however is that the angular distribution depends on the production process and can be quite different even for similar processes, for example when comparing the production of $G+\gamma$ to $G + {\rm jet}$ production. It is therefore not possible to conclusively exclude a nonzero spin of $\phi_a^0$ just using this observable, or in other words, just using the three-photon final state.\footnote{In principle, however, one can obtain a lower bound on the spin by finding the highest nonvanishing cosine or sine mode.}

An alternative spin determination method was proposed in Ref.~\cite{Atwood10} for the analysis of associated KK-graviton--photon production with the graviton decaying into a photon pair. They find that the angular distribution of one photon from this pair in the rest frame of the decaying graviton is sensitive to the graviton spin, but again the details of this distribution also depend on the production process.

\vspace{\bigskipamount}

\begin{figure}
\begin{center}
\epsfig{figure=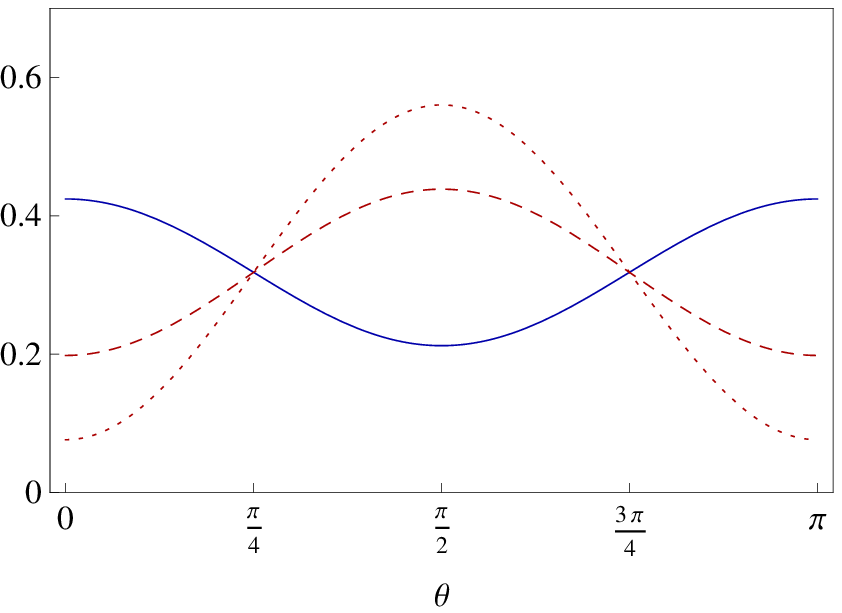, width=8.5cm}%
\end{center}
\vspace{-1.5em}
\mycaption{Normalized distribution of the angle $\theta$ between the photon and the lepton in a decay chain $X \rightarrow Z \gamma \rightarrow \gamma \ell \bar\ell$. The blue solid line corresponds to the pseudoscalar $\phi_a^0$ while the dashed and dotted red lines show the distributions for a spin two graviton with mass $M_G/M_Z = 2$ and $M_G/M_Z=4$ respectively. }
\label{fig:zll}
\end{figure}

A more reliable measurement of the $\phi_a^0$ spin that is independent of the production channel can be obtained by looking at different decay channels. One promising channel is the decay $\phi_a^0 \rightarrow \gamma Z$ with a subsequent decay of the $Z$ boson into charged leptons. The $\phi_a^0 \rightarrow \gamma Z$ branching fraction is about 30\% over most of the relevant parameter space. The angle $\theta$ between the photon and either of the leptons is sensitive to the polarization orientation of the $Z$ boson and thus to the spin of the parent particle. For a pseudoscalar resonance decaying through the interaction (\ref{eq:wzw}) we obtain
\begin{align}
	\frac{d \Gamma}{d \cos \theta} & \propto (1 + \cos^2 \theta ) \,.
\end{align}
For comparison, the $\theta$ dependence for the decay of a spin-two Kaluza-Klein graviton, $G \rightarrow Z \gamma \rightarrow \gamma \ell^+ \ell^-$ \cite{KKG}, is 
\begin{align}
	\frac{d \Gamma}{d \cos \theta} & \propto 6 M_G^2 \sin^2\theta + 7 M_Z^2 (1 + \cos^2\theta )\,,
\end{align}
which is clearly distinguishable from the spin-0 case. The normalized angular distributions for the spin-zero case and for the spin-two case with $M_G/M_Z = 2$ and $M_G/M_Z = 4$ are shown in figure \ref{fig:zll}. 

For completeness, we point out that a heavy vector boson $Z'$ 
can also decay to the $\gamma Z$ final state through an anomaly interaction \cite{perelqi}, all the while the $\gamma\gamma$ channel is forbidden for a spin-one particle.
The resulting $\theta$ distribution is given by
\begin{align}
	\frac{d \Gamma}{d \cos \theta} & \propto M_{Z'}^2 \sin^2\theta + 
	M_Z^2 [c_c (1 + \cos^2\theta ) + c_s \sin^2\theta] +
	{\cal O}(M_Z^4/M_{Z'}^2) \,,
\end{align}
where $c_{c,s}$ are ${\cal O}(1)$ constants which depend on the details of the Chern-Simons and Wess-Zumino-Witten terms that mediate the decay.
For $M_{Z'} \gg M_Z$, this distribution peaks at $\theta=\pi/2$, similar to the KK-graviton case, and thus can be clearly distinguished from a spin-zero parent particle. For $M_{Z'} \approx M_Z/2$, however, the $\theta$ dependence of the $Z'$ vector boson could look similar to the spin-0 or the spin-2 case, depending on the values of $c_{c,s}$. However, as mentioned above, a distinction is always possible by observing the $\gamma\gamma$ decay mode, which is forbidden for a vector boson.

The CP properties of the scalar resonance can not be determined from the $Z\gamma$ channel, since a parity-even scalar that decays through the effective coupling $\phi Z_{\mu\nu} A^{\mu\nu}$ displays the same $(1+ \cos^2 \theta)$ behavior as the pseudoscalar. 

Full information about the particle properties can be obtained from the so called golden channel decay \cite{Choi:2002jk}
\begin{align}
    \phi^0_a \longrightarrow ZZ^{(*)} \longrightarrow \ell^+ \ell^- \ell^+ \ell^- \,,
\end{align}
where $\ell$ denotes charged leptons, in particular electrons and muons. Detailed studies on how to extract spin information using these channels have been recently performed in Refs.~\cite{Melnikov10,Lykken10}, which show that a small number of events (20--30) is enough to obtain a $3 \sigma$ discrimination between different spin and parity hypotheses. Due to the small branching fraction of the $Z$ boson into leptons this channel can only be used for spin determination if the $\phi_a$ are light, below $m_a = 300 \gev$, even for the full LHC design luminosity of 300~fb$^{-1}$. 
\subsection{Extracting Model Information}
An important feature of the anomaly-mediated decays of the scalar triplet is that the branching fractions of $\phi^0_a$ and $\phi_a^\pm$ are completely fixed by the global symmetry structure. If these particles are found in the multi-photon channel the next task will be to look for the possible decays of $\phi^0_a$ into $Z\gamma$ and $ZZ$ pairs as well as for the $ZW^\pm$ decay mode of the $\phi^\pm_a$. The measured branching fractions will then unambiguously determine whether the decays are mediated by the WZW term and thus reveal information about the symmetry structure of the model. 

If the mass splitting between the charged and neutral components can be measured, for example indirectly by observing the $4\gamma +X$ decay modes, more information can be obtained. Since the splitting is purely due to radiative corrections, a mass splitting larger than the one generated by standard model gauge interactions, see eq.~\eqref{eq:masssplit}, hints towards the existence of additional global symmetry breaking operators, $e.\,g.$ additional gauge bosons or new fermions, as in the case of little Higgs models. 
\section{Conclusions}
\label{pheno:end}
In this paper we discuss multi-photon signals from pseudoscalar electroweak triplets $\phi_a^{0,\pm}$ and the prospects for the LHC to detect them. These triplet fields appear in a general class of composite extensions of the standard model, in particular in little Higgs models and in models where parts of the fermion sector transforms vector-like under the electroweak SU(2) gauge symmetry. 
The multi-photon signal is therefore an important channel for detection or exclusion of these models.

In these models, the leading-order effective Lagrangian for the pseudoscalars is symmetric under some parity, so that their decays are only induced loop-level Wess-Zumino-Witten terms that lead to anomalous breaking of the parity. As a result, the decay channels and branching fractions are fully determined by the global symmetry structure. In particular we find that the neutral component has a large branching fraction into photon pairs even for masses well above the $ZZ$ threshold. Due to the parity symmetry, at the LHC the triplets are mostly produced in pairs, with a sizable cross section for masses up to $m_a\sim 600 \gev$. 

The two major sources of backgrounds for $3\gamma$ signals at the LHC are direct three photon production an $2\gamma + n \;{\rm jet}$ production where one of the jets is misidentified as a photon. We generated Monte-Carlo events with matrix-element matching and used the fast detector simulation PGS4 to estimate the backgrounds and in particular the photon fake rate. and to account for detector efficiencies and energy smearing. We find that the background can be reduced significantly with a set of simple cuts, so that a $5\sigma$ discovery is possible at the 14 TeV LHC with $30\;{\rm fb}^{-1}$ for pseudoscalar masses up to about 500~GeV. For the current 7 TeV run of the LHC a discovery is still possible if the $\phi_a$ are lighter than 250~GeV. 

Since this signal relies only on a clean signature involving photons, and possibly leptons from $W$-boson decays, we expect that the masses of the neutral $\phi_a^0$ and charged $\phi_a^\pm$ can be measured with high accuracy from the di-photon invariant mass peak and the kinematic endpoint of the lepton-photon invariant mass distribution, respectively. In principle, the spin of the neutral pseudoscalar $\phi_a^0$ can be determined from the angular distribution of the two-photon final state but, depending on the mass of $\phi_a^0$, the spin effect can be quite small and difficult to measure experimentally.

Other decay modes of the triplets, in particular into $ZZ$ or $Z \gamma$ final states, offer additional possibilities to obtain information about the nature of the observed particles. We have shown that the decay of $\phi_a^0$ into $Z \gamma$ followed by the leptonic decay of the $Z$ boson can be used to distinguish it from a resonance with spin two, and possibly also spin one.
In the $ZZ$ channel, the fully leptonic decay of both $Z$ bosons can be used to uniquely determine the spin and the parity of the $\phi_a^0$. The branching fractions for the different channels yield information about the global symmetry structure of the model.

Finally, it is interesting to note that the current limit for the $3\gamma+X$ signal from the D\O\ experiment at the Tevatron was obtained with only $0.83 \;{\rm fb}^{-1}$ of data, and translates into a bound of $m_a>150\gev$ for our model. This bound could be improved drastically by using the full data sets that are now available at D\O\ and CDF.

\section*{Acknowledgements}

We would like to thank D.~Wyler for stimulating discussions, R.~Frederix for help with MadGraph and A.~Belyaev for help with CalcHEP. This project was supported in part by the Schweizer Nationalfonds, the National Science Foundation under grant PHY-0854782,  and by the U.S.\ Department of Energy, Division of High Energy Physics, under Contract DE-AC02-06CH11357 and  DE-FG02-84ER40173.



\begin{thebibliography}{99}
\frenchspacing


\bibitem{tcref}
  C.~T.~Hill and E.~H.~Simmons,
  Phys.\ Rept.\  {\bf 381}, 235 (2003)
  [Erratum-ibid.\  {\bf 390}, 553 (2004)].
  
\bibitem{schmaltz}
  M.~Schmaltz and D.~Tucker-Smith,
  Ann.\ Rev.\ Nucl.\ Part.\ Sci.\  {\bf 55}, 229 (2005).



\bibitem{vec1}
  C.~Kilic, T.~Okui and R.~Sundrum,
  JHEP {\bf 1002} (2010) 018.
  
  
\bibitem{WZW}
  J.~Wess and B.~Zumino,
  Phys.\ Lett.\  B {\bf 37}, 95 (1971);\\
  E.~Witten,
  Nucl.\ Phys.\  B {\bf 223}, 422 (1983);\\
  O.~Kaymakcalan, S.~Rajeev and J.~Schechter,
  Phys.\ Rev.\  D {\bf 30}, 594 (1984).
  

\bibitem{LHT}
  H.~C.~Cheng and I.~Low,
  %
  JHEP {\bf 0408}, 061 (2004).
 

\bibitem{hillsq}
  C.~T.~Hill and R.~J.~Hill,
  Phys.\ Rev.\  D {\bf 75}, 115009 (2007);\\
  C.~T.~Hill and R.~J.~Hill,
  Phys.\ Rev.\  D {\bf 76}, 115014 (2007).

\bibitem{lhtwzw}
  V.~Barger, W.~Y.~Keung and Y.~Gao,
  Phys.\ Lett.\  B {\bf 655}, 228 (2007);\\
  A.~Freitas, P.~Schwaller and D.~Wyler,
  JHEP {\bf 0809}, 013 (2008).

\bibitem{Bai1}
  Y.~Bai and A.~Martin,
  Phys.\ Lett.\  B {\bf 693}, 292 (2010).
  
  \bibitem{Krohn:2008ye}
  D.~Krohn and I.~Yavin,
  JHEP {\bf 0806}, 092 (2008).
  
   \bibitem{mmx}
  A.~Freitas, P.~Schwaller and D.~Wyler,
  JHEP {\bf 0912}, 027 (2009); \\
  P.~Schwaller, 
  AIP Conf.\ Proc.\  {\bf 1200}, 615 (2010).
  
\bibitem{vec2}
  C.~Kilic and T.~Okui,
  JHEP {\bf 1004}, 128 (2010).
  

\bibitem{Bai2}
  Y.~Bai and R.~J.~Hill,
  arXiv:1005.0008 [hep-ph].
  
  

\bibitem{Cirelli}
  M.~Cirelli, N.~Fornengo and A.~Strumia,
  Nucl.\ Phys.\  B {\bf 753}, 178 (2006).
 
\bibitem{proc09}  
  A.~Freitas, P.~Schwaller and D.~Wyler,
  Nucl.\ Phys.\ Proc.\ Suppl.\  {\bf 200-202}, 169 (2010);
  arXiv:0912.3647 [hep-ph].

 \bibitem{comphep}
  E.~Boos {\it et al.}  [CompHEP Collaboration],
  Nucl.\ Instrum.\ Meth.\ A {\bf 534}, 250 (2004).

\bibitem{pythia}
  T.~Sj\"ostrand, S.~Mrenna and P.~Z.~Skands,
  JHEP {\bf 0605}, 026 (2006).
  
\bibitem{pgs}
J.~Conway, {\tt  http://www.physics.ucdavis.edu/\~{}conway/research/software/pgs/\linebreak[0]pgs4-general.htm}.


\bibitem{dzero0}
    The~D\O~Collaboration, D\O\ Note 5067-CONF.
    
\bibitem{Abazov:2010xh}
  V.~M.~Abazov {\it et al.}  [The D\O\ Collaboration], 
  arXiv:1004.1826 [hep-ex]. 

\bibitem{cdfpublic}
The CDF Collaboration, public note, {\tt http://www-cdf.fnal.gov/physics/exotic/\linebreak[0]r2a/20100513.graviton.diphoton/index.html}.

\bibitem{madgraph}
  J.~Alwall {\it et al.},
  JHEP {\bf 0709}, 028 (2007).

\bibitem{2phot}
  C.~Bal\'azs, P.~M.~Nadolsky, C.~Schmidt and C.~P.~Yuan,
  Phys.\ Lett.\  B {\bf 489}, 157 (2000);\\
  Z.~Bern, L.~J.~Dixon and C.~Schmidt,
  Phys.\ Rev.\  D {\bf 66}, 074018 (2002).
  
\bibitem{madmatch}
See $e.\,g.$ {\tt http://cp3wks05.fynu.ucl.ac.be/twiki/bin/view/Main/IntroMatching}.
    
\bibitem{cms_photon}
  M.~Pieri, K.~Armour and J.~G.~Branson, CERN-CMS-NOTE-2006-007.


\bibitem{atl_photon}
  F.~Derue, ATL-COM-PHYS-2005-046.



\bibitem{Murayama:2009jz}
  M.~R.~Buckley, H.~Murayama, W.~Klemm and V.~Rentala,
  Phys.\ Rev.\  D {\bf 78} (2008) 014028; \\
  H.~Murayama and V.~Rentala,
  arXiv:0904.4561 [hep-ph].
  
\bibitem{Hagiwara:2008jb}
  K.~Hagiwara, J.~Kanzaki, Q.~Li and K.~Mawatari,
  Eur.\ Phys.\ J.\  C {\bf 56}, 435 (2008).
 

\bibitem{Atwood10}
  D.~Atwood and S.~K.~Gupta,
  arXiv:1006.4370 [hep-ph].

\bibitem{KKG}  
  B.~C.~Allanach, J.~P.~Skittrall and K.~Sridhar,
  JHEP {\bf 0711}, 089 (2007);\\
  J.~P.~Skittrall,
  Eur.\ Phys.\ J.\  C {\bf 60}, 291 (2009).
  
\bibitem{perelqi}
  M.~Perelstein and Y.~H.~Qi,
  Phys.\ Rev.\  D {\bf 82}, 015004 (2010).
       
\bibitem{Choi:2002jk}
  S.~Y.~Choi, D.~J.~Miller, M.~M.~M\"uhlleitner and P.~M.~Zerwas,
  Phys.\ Lett.\  B {\bf 553}, 61 (2003).


\bibitem{Melnikov10}
  Y.~Gao, A.~V.~Gritsan, Z.~Guo, K.~Melnikov, M.~Schulze and N.~V.~Tran,
  Phys.\ Rev.\  D {\bf 81}, 075022 (2010).


\bibitem{Lykken10}
  A.~De Rujula, J.~Lykken, M.~Pierini, C.~Rogan and M.~Spiropulu,
  Phys.\ Rev.\  D {\bf 82}, 013003 (2010).



\end{thebibliography}
\end{document}